\documentstyle[prl,floats,aps,psfig,twocolumn]{revtex}
\def\etal{{\sl et al.}}                 %et al. - no preceeding comma
\def\Journal#1#2#3#4{{#1} {\bf #2}, #3 (#4)}
\def\NPB{{Nucl. Phys.} B}
\def\PLB{{Phys. Lett.}  B}
\def\PRL{Phys. Rev. Lett.}
\def\APP{Astro. Part. Phys.}
\def\PRD{{Phys. Rev.} D}
\def\PRC{{Phys. Rev.} C}

\def\REM{Rev. Mod. Phys.}

                        \def\be{\begin{equation}}
\def\ee{\end{equation}}                     \def\bea{\begin{eqnarray}}
\def\eea{\end{eqnarray}} 

\begin{document}
%
%\preprint{Version 12}
\draft
\title{A study  of the solar neutrino survival probability}

\date{\today}

% LIST_OF_AUTHORS.TEX                 10/30/97
%
\author{C.~M.~Bhat$^a$,  P.~C.~Bhat$^a$, M.~Paterno$^b$ and  H.~B.~Prosper$^c$ }  

\address{$^a$ Fermi National
Accelerator  Laboratory,
P.O. Box  500,  Batavia, IL~60510 \\
$^b$ Department  of Physics and Astronomy, University   of  Rochester,  Rochester, New York 14627\\
$^c$ Department  of Physics, Florida  State University, Tallahassee, Florida 32306} 

\maketitle
\begin{abstract}
We present  a study of  recent  solar neutrino  data using a  Bayesian
method.    Assuming   that   only   $\nu_e$   are   observed   in  the
Super-Kamiokande  experiment our results  show a  marked supression of
the survival probability at about 1~MeV, in  good agreement with $\chi
^2$-based   analyses.   When      the detection  of    $\nu_{\mu}$  by
Super-Kamiokande is    taken   into  account,   assuming $\nu_e$    to
$\nu_{\mu}$ oscillations, we find the  largest suppression in survival
probability at about 8.5~MeV.
\end{abstract}


%\nopagebreak
%\nopagebreak

%\twocolumn

One of the most  intriguing problems of the past  two decades has been
the observation of a deficit of  neutrinos of solar origin as compared
to  the  predictions of  standard solar 
models~\cite{bp95,BahcallBook,ssmTL}.
Many attempts have  been made to explain this  discrepancy either as a
consequence of astrophysical processes or new physics such as neutrino
oscillations in  vacuum~\cite{gribov} or in matter~\cite{wolfenstein}.
The   astrophysical         solutions         have       not      been
successful~\cite{astro}.   However, the  solutions    that invoke  new
physics provide excellent descriptions of    the solar neutrino   
data~\cite{hata,lui}.  In this paper we answer the following
question: how well do we know the
neutrino survival
probability as a function of the neutrino energy? The answer is
relevant because it tells us in 
what way, and to what degree, the solar 
neutrino data constrain
the models that seek to explain the neutrino deficit.
We answer the question using a Bayesian method. Our perspective here
is broader than that of Ref.~\cite{gates} in which a Bayesian
method was used to analyze the MSW model.

In our analysis  we assume  that the solar
neutrino    spectrum is  that    predicted    by the standard    solar
models~\cite{bp95,BahcallBook,bb8}.  However,  it    is known
that the spectrum is insensitive to the  details of these models~\cite{bnmodels}. 
The experimental  data  are from     Homestake  (Cl)~\cite{slrCL}, SAGE  
(Ga)~\cite{slrGa1},
GALLEX        (Ga)~\cite{slrGa2}            and       Super-Kamiokande
(H$_2$O)~\cite{slrSK,Fogli}. These results  together with the predictions of
the standard solar model  of Bahcall and Pinsonneault~\cite{bp95}  are
shown  in Table   I.  Our method  of  analysis  does  not require  the
imposition of the solar luminosity constraint.  In accordance with our
minimalist approach we choose not to impose it.

%%\section{THEORY OF BAYESIAN ANALYSIS}

Bayes' theorem,
  $P(H|D,I) =$ ${\cal L}(D|H,I)$ $P(H|I)$
           $/ \int_{_{H}} {\cal L}(D|H,I)$ $P(H|I)$,
gives a prescription for calculating the posterior
probability $P(H|D,I)$ of an hypothesis $H$, given measured quantities
$D$ and prior information $I$;  
$\cal L$ is the likelihood function assigned to $D$ and $P(H|I)$
is the prior probability assigned to $H$.  The integration in the
denominator is over all hypotheses of interest.

The solar neutrino rate $S_i$ for the Chlorine and Gallium experiments
is given by,
\begin{equation}
S_i = {\sum_j} \Phi_j \int_{E_{th_i}}\sigma_i(E_{\nu}) \phi_j(E_{\nu})
{\cal P}(E_{\nu})dE_{\nu},
\label{eq:radio}
\end{equation}
where $\Phi_j$ is the total flux from neutrino source $j$, $\phi_j$ is
the corresponding normalized neutrino energy spectrum, $\sigma_i$ is
the cross-section for the $i\mbox{th}$ experiment, $E_{th_i}$ is its
threshold energy (see Table I) and ${\cal P}(E_{\nu})$ is the neutrino
survival probability.

For the Super-Kamiokande experiment we use their reported
measurement of the electron recoil spectrum produced by the $^8B$
neutrinos\cite{Fogli}, that spans the range 6.5 to 20~MeV.
If the $\nu_e$ deficit is
caused by $\nu_e$ oscillations to $\nu_{\mu (\tau)}$ then one must
take account of the fact that Super-Kamiokande is sensitive to (but
does not distinguish between) all flavors of neutrino.  On the other
hand, if the $\nu_e$ disappear through a mechanism that does not
result in other detectable particles, for example by oscillating into
sterile neutrinos, then the measured rate is to be ascribed to the
$\nu_e$ flux only.  We consider both possibilities.

The {\em measured} electron recoil spectrum $N(T)$ is given by
\begin{eqnarray}
N(T) & = & N_0 \int_0^{T^{max\prime}(E_{\nu}^{max})} dT^{\prime}
R(T|T^{\prime}) \\ \nonumber & \times &
\int_{E_{\nu}^{min}(T^{\prime})}^{E_{\nu}^{max}} dE_{\nu}
\phi_B(E_{\nu})[{\cal P}(E_{\nu})\sigma_e(T^{\prime},E_{\nu}) \\
\nonumber & + & (1-{\cal
P}(E_{\nu}))\sigma_{\mu}(T^{\prime},E_{\nu})],
\label{eq:superk}
\end{eqnarray}
where $R(T|T^{\prime})$ is the Super-Kamiokande resolution function
(which can be approximated by a Gaussian with mean $T^{\prime}$ and
standard deviation $1.5\sqrt(T^{\prime}/10\mbox{MeV})$\cite{Fogli}),
$T = E_e - m_e$ and $T^{\prime}$ are the measured and true electron
kinetic energies, respectively, with $E_e$ and $m_e$ the
electron energy and mass. The quantity $\phi_B$ is the normalized  neutrino 
energy spectrum from the $^8B$ reaction, and  
$\sigma_e$ and $\sigma_{\mu}$  are the
$\nu_e$ and $\nu_{\mu}$ differential electron scattering
cross-sections\cite{Bcross}, respectively.  Given a neutrino energy $E_{\nu}$ the
electron can assume a maximum kinetic energy of $T^{max\prime}(E_{\nu}) = 2
E_{\nu}^2/(2 E_{\nu}+m_e)$, while the minimum neutrino energy for a
fixed $T^{\prime}$ is given by $E_{\nu}^{min}(T^{\prime}) =
[T^{\prime}+\sqrt(T^{\prime} (T^{\prime}+2 m_e))]/2$;  $E_{\nu}^{max}$
is the maximum neutrino energy, which we take to be 20~MeV.  The
constant $N_0$ is a normalization factor that depends on which units
are used for the event rate.
\begin{table}% [t]
\caption{Measured and predicted solar  neutrino capture rates~[16], in units of SNU
(=$10^{-36}$  captures/atom~s$^{-1}$), for the radiochemical
experiments, and neutrino fluxes, in units of
$10^6$ cm$^{-2}$ s$^{-1}$, for  Super-Kamiokande. 
} 
\begin{center}
\begin{tabular}{cccl}  
 Experiment  & E$_{th}$   & \multicolumn{2}{c}{Flux Rates} \\ 
\cline{3-4}  & (MeV)                     & Measured      & Predicted\\   \hline

Homestake    & 0.814   & 2.54$\pm$0.16$\pm$0.14 & 9.30$\pm$1.30 \\ 

($\nu_e$+$^{37}$Cl$\rightarrow   e^-$+$^{37}$Ar)    &    & & \\ \hline
 
GALLEX       &  0.233 & 76.2$\pm$6.5$\pm$5 & 137$\pm$8 \\ 
SAGE         &''                                           & 73$\pm$8.5$_{-6.9}^{+5.2}$ &~~''   \\
($\nu_e$+$^{71}$Ga $\rightarrow e^-$+$^{71}$Ge)               &   & & \\ \hline

Super-Kamiokande       &   6.5   & 2.44$\pm$0.06$_{-0.09}^{+0.25}$ & 6.62$_{-1.12}^{+0.93}$ \\ 
($\nu_e$+$e^- \rightarrow  e^-$+$\nu_e$)  &                              & & \\ 
\end{tabular} \end{center} 
\vspace{-0.2in}
\end{table}
In writing Eq.~(2) we have assumed $\nu_e$ to
$\nu_{\mu}$ oscillations.  If the measured flux is due to $\nu_e$ only
the recoil spectrum is given by Eq.~(2) with the term
proportional to $\sigma_{\mu}$ omitted. The event rate $S_i$ in the
$i$th electron energy bin is simply $N(T)$ integrated over that
 bin.

\begin{figure}
\centerline{\psfig{figure=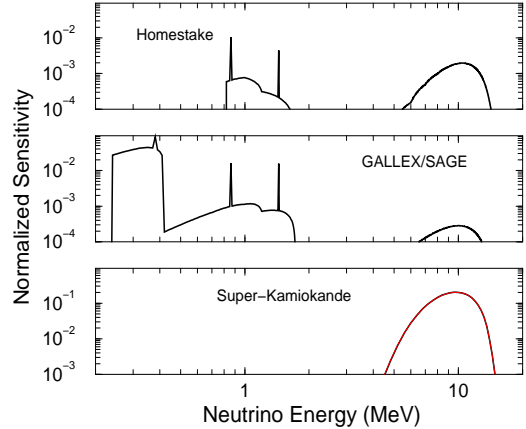,width=3.in,height=2.6in }}
\caption{Spectral sensitivity as a function of the neutrino
energy.}
\label{fig:sensitivity}
\vspace{-.1in}
\end{figure}

We note that each experiment is sensitive to different parts of the
neutrino energy spectrum. This is evident from the (normalized) plots
of $\sigma(E_{\nu}) {\sum_j} \Phi_j \phi_j(E_{\nu})$ 
(or $\phi_B(E_{\nu})$ $\times$ 
$\int_{6.5-m_e}^{20-m_e} dT$ $\int_{0}^{T(E_{\nu})}$
$dT^{\prime} R(T|T^{\prime})$ $\sigma_e(T^{\prime},E_{\nu})$
for
Super-Kamiokande)
 shown in Fig.~\ref{fig:sensitivity}.  We also note
the existence of regions where the spectral sensitivity is
essentially zero.

 The likelihood function ${\cal L}(D|H,I)$ 
is assumed to be of Gaussian form $g(D|S,\Sigma)$,
where $D \equiv (D_1,\ldots,D_{18})$ represents the 18 data---2 rates
from the radiochemical experiments plus 16 rates from the binned
Super-Kamiokande electron recoil spectrum, $\Sigma $ is the
$18\times18$ error matrix for the experimental data and $S \equiv
(S_1,\ldots,S_{18})$ represents the predicted rates. The error matrix
is deduced from the data given in Ref.~\cite{Fogli}.  We take the prior
probability  to be constant.

We extract the survival probability using two different methods: {\it
binned} and {\it parametric}.  In the {\it binned} method we divide
the neutrino energy spectrum into 12 bins between 0.2 and 20 MeV,
chosen so that the survival probability within each bin is
approximately constant.  With a minor algebraic manipulation of
Eq.~(1), the neutrino rates $S_1$ and $S_2$ for the
Chlorine and Gallium experiments, respectively, reduce to a weighted
sum of ${\cal P}_k$ (the survival probability in bin k): 
\begin{eqnarray}
S_i & \approx & \sum_k {\cal P}_k \sum_j \Phi_j
\int_{E_k}^{E_{k+1}} \sigma_i(E_{\nu}) \phi_j(E_{\nu}) dE_{\nu},
\end{eqnarray}
while for Super-Kamiokande, assuming $\nu_e$ to $\nu_{\mu}$
oscillations, the neutrino rate is expressed as sum of 
 the sixteen  spectral values $S_3$ to $S_{18}$, each containing two terms: 
one with $\sigma_{\mu}$ and the second with
the difference between $\sigma_{e}$ and $\sigma_{\mu}$ 
(see Eq.~(2)).
If electron neutrinos are the only detectable particles then one 
gets a similar expression with the $\sigma_{\mu}$ terms omitted.

\begin{figure}
\centerline{\psfig{figure=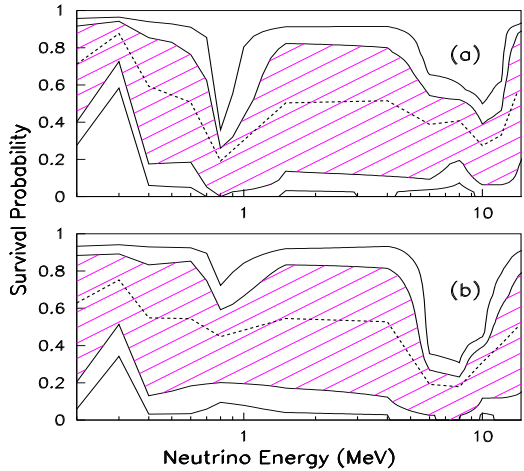,width=3.0in,height=2.6in}}
\caption{
Our best estimate of the survival probability (dark dashed line) {\it vs} 
neutrino energy for the  binned  method
assuming   (a)~the neutrino flux consists of $\nu_e $ only  and (b)~ $\nu_e $ 
plus $\nu_{\mu}$, due to oscillations.
The 68.3\% (shaded area) and 90\% (solid line) 
confidence intervals, about our best estimate, are also shown. 
}
\label{fig:binnede}
\vspace{-.1in}
\end{figure}

From Bayes' theorem we compute the posterior probability
$P({\cal P}|D,I)$, that is, the probability of the hypothesis $H$ that
the set of parameters ${\cal P} \equiv ({\cal P}_1,\ldots,{\cal
P}_{12})$ have specified values.  For the Gallium rate we use the
weighted average of the SAGE and GALLEX results.  The posterior
probability $P({\cal P}_k|D,I)$ for each ${\cal P}_k$ is obtained by
marginalizing (that is, integrating) $P({\cal P}|D,I)$ over the
remaining ${\cal P}_k$. We use the mean of $P({\cal P}_k|D,I)$
as our
best estimate of ${\cal P}_k$.
The results, assuming that the measured rate is
produced by $\nu_e$ only, are shown in Fig.~\ref{fig:binnede}(a).  
We see that in the first two bins, that span
the $pp$ neutrino spectrum, the survival
probability is close to unity. There is a marked suppression in the
bin (0.8-1.5~MeV) that contains the main $^7Be$ line and a moderate
suppression in the $^8B$ spectrum (bins spanning 6-12~MeV).  These
results agree with previous analyses~\cite{hata,lui}.
But our result provides additional information, namely: the 
precision with
which the survival probability is known---as a function of the
neutrino energy, irrespective of the precise origin of the neutrino
deficit. Moreover, since our binned method makes no assumption about
the form of the survival probability it makes no unwarranted
inferences. Where there is little spectral sensitivity (in the
intervals 0.4-0.8 MeV and 1.5-4.5 MeV) our method infers correctly
that little information can be extracted.

If, however, we assume oscillations into active neutrinos
($\nu_{\mu}$) the inferred form of the survival probability
changes. Figure~2(b) shows  
the extracted survival probability for such a case. 
Here,
we see a marked suppression at a higher neutrino
energy ({\it i.e.,} at around 8.5~MeV).

In the parametric method,
which is  motivated by the conclusions of previous
analyses~\cite{hata,lui,parke},
we write
the survival probability as a sum of
two finite Fourier series:
\begin{eqnarray}
{\cal P}(E_{\nu}|a) & = 
& \sum_{r=0}^{7} a_{r+1} \mbox{cos}(r\pi E_{\nu}/L_1) \\ \nonumber
& / & (1 + \mbox{exp}[(E_{\nu}-L_1)/b])  \\ \nonumber 
& + & \sum_{r=0}^{3} a_{r+9} \mbox{cos}(r \pi E_{\nu}/L_2).
\label{eq:parametric}
\end{eqnarray}
The first
term in Eq.~(4) is defined in the
interval 0.0 to $L_1$ MeV---and suppressed
beyond $L_1$ by the exponential, while the second term 
covers the interval 0.0 to $L_2$ MeV. 
We divide the function in this way so that it
can accomodate a survival probability 
that varies rapidly in the interval 0.0
to $L_1$.
Holding the
parameters $L_1$, $L_2$ and $b$ fixed at 1.0, 15.0 and 0.1~MeV,
respectively, we compute the posterior
probability $P(a|D,I)$ for the parameters $a \equiv
(a_1,\ldots,a_{12})$.  The data are the same as those used for
the binned method; but instead of a linear sum in ${\cal P}_k$ 
we have a linear sum in the parameters $a$.

The theoretical uncertainties,
which so far we have neglected, can be
incorporated by treating 
the fluxes $\Phi \equiv (\Phi_1,\ldots)$ as
parameters with an associated prior probability, $P(\Phi|I)$, that
encodes the flux predictions and whatever correlations exist amongst
them.  
We represent our knowledge of the fluxes
by a multivariate Gaussian prior probability
$P(\Phi|I) =$ $g(\Phi|\Phi^0,\Sigma_{\Phi})$,
where $\Phi^0 \equiv (\Phi_1^0,\ldots,\Phi_8^0)$ is the vector of flux
predictions and $\Sigma_{\Phi}$ is the corresponding error 
matrix\cite{bahcall97}.
For simplicity we neglect correlations; hence
 $\Sigma_{\Phi}$ is diagonal.

\begin{figure}
\centerline{\psfig{figure=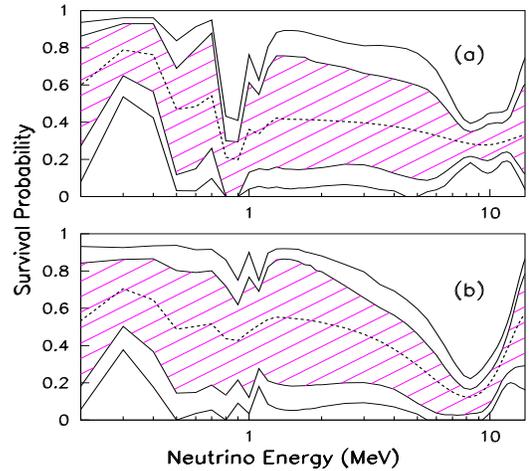,width=3.0in,height=2.6in}}
\caption{
Survival probability {\it vs} neutrino energy for the   parametric  method
assuming (a)~the  neutrino flux consists of $\nu_e $ only and 
(b)~$\nu_e $ to $\nu_{\mu}$. 
All uncertainties  are included.
See Fig.~\ref{fig:binnede} for other details.}
\label{fig:parme}
\vspace{-.1in}
\end{figure}

The posterior probability (using 
 a constant prior probability for the parameters $a$ is now given
by
$P(a,\Phi|D,I) =$ ${\cal L}(D|a,\Phi,I)$ $P(\Phi|I)$
$/\int_{a,\Phi}{\cal L}(D|a,\Phi,I)$ $P(\Phi|I)$,
which when marginalized with respect to $\Phi$ gives
$P(a|D,I)$.
The  survival probability is estimated using 
$P(E_{\nu}|D,I) =$ $\int_{a} {\cal P}(E_{\nu}|a)$ $P(a|D,I)$.
Figure~\ref{fig:parme}(a) shows the survival probability and its
uncertainty, which now includes both the experimental and theoretical
errors.  Again, the general form of the survival probability obtained,
assuming the neutrinos detected consist of  $\nu_e $ only, agrees with the
inferences from  previous analyses and our binned method.  But
again, unlike previous analyses we have detailed information about the
survival probability and its uncertainties as a function of the
neutrino energy.  
Figure~\ref{fig:parme}(b) shows our results, for the
survival probability, assuming oscillations into active
neutrinos.

One useful check of the reliability of both calculations, and
the flexibility of the function in Eq.~(4), is
to set the measurements $D$ equal to their predicted values and to
verify that the extracted survival probability is unity in the energy
ranges where data exist and is 0.5 with an error of $\pm 0.3$ where
there are no data.  Both calculations have been verified successfully
in this way. 
We note the general agreement between the binned and
parametric
methods in the energy regions with adequate spectral
sensitivity. 

\begin{figure}
\centerline{\psfig{figure=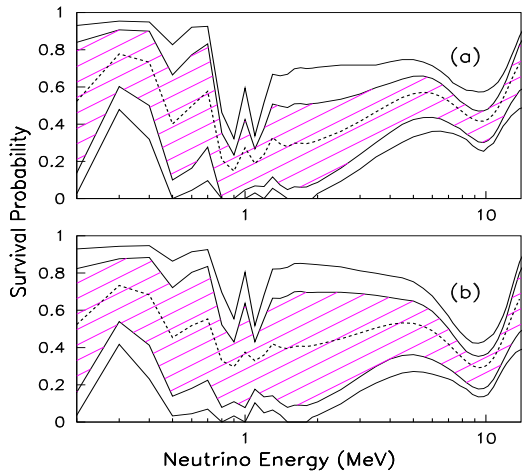,width=3.0in,height=2.6in}}
\caption{
Survival probability {\it vs} neutrino energy for the   parametric  method
using 1998 data and solar model[21]. All uncertainties  are included.
See Fig.~3 for other details.}
\label{fig:par98}
\vspace{-.1in}
\end{figure}

The calculations presented here 
are based on data from 1997. 
Our 
preliminary calculations using the 
data presented at the recent Neutrino '98 conference\cite{neutrino98}, 
and the 1998 Bahcall and Pinsonneault predictions\cite{bahcall98}, 
yield similar conclusions, as illustrated in
Fig.~\ref{fig:par98}.
The Bayesian method 
offers a well-founded way to
  compute the relative probabilities of different models
of the survival probability.
We plan to provide such
  calculations in a longer paper.

In summary, we have used  recent solar neutrino data and  
standard solar model predictions to extract the neutrino survival
probability and its uncertainty as a function of the neutrino energy
under two broad alternate assumptions: 1) the measured flux is
comprised of $\nu_e$ only or 2) it is a mixture of active
neutrinos  arising from $\nu_e$ oscillations.
Under assumption 1), we find that the survival
probability is most precisely determined around 0.3~MeV and has a
value of 0.79$\pm$0.13 (and 0.28$\pm$0.03 at 8.3~MeV) with
experimental and theoretical uncertainties included.  However, the
uncertainties elsewhere are considerably larger. 
Our results suggest that the data could
accomodate a wider range of models than hitherto have been considered.
In particular, the fact that the MSW model provides a good description
of the data is clearly a strong point in its favor; but it is at
present not a decisive one. We therefore eagerly await further results
from Super-Kamiokande and the first results from SNO\cite{sno}.

%\acknowledgments 
The authors would like to thank John Bahcall for
providing the latest theoretical information   used in the 
present analysis and for useful suggestions and Robert Svoboda for providing  the 
latest Super Kamiokande data. We acknowledge
the support of the U.S. Department of Energy and U.S. National Science Foundation.
Fermi National Accelerator Laboratory is operated by the Universities
Research Association, under contract with the U.S. Department of Energy.

\sloppy

\end{document}